\documentstyle[11pt,newpasp,twoside,epsf]{article}
\def\edcomment#1{\iffalse\marginpar{\raggedright\sl#1\/}\else\relax\fi}
\reversemarginpar

\begin{document}
\title{Formation and Migration of Trans-Neptunian Objects}

\author{S.I. Ipatov}
\affil{NRC/NAS Senior Research Associate at
 NASA/GSFC, Greenbelt, 20771, USA; Institute of Applied Mathematics, 
Miusskaya sq. 4, Moscow 125047, Russia}

\begin{abstract}
Some large trans-Neptunian objects 
could be formed by the compression of rarefied dust condensations, but not by the 
accumulation of smaller planetesimals. A considerable portion of near-Earth objects 
could have come from the trans-Neptunian region. Our runs of the evolution of thousands 
orbits of Jupiter-family comets under the 
gravitational influence of planets showed that former Jupiter-family comets 
collide with the terrestrial planets mostly from orbits with aphelia located deep
inside Jupiter's orbit. 

\end{abstract}

\section{Formation of Tran-Neptunian Objects}

Many scientists considered that large trans-Neptunian objects (TNOs) and asteroids
were formed by accumulation of smaller (e.g., 1-km) planetesimals.  Such 
process of accumulation of TNOs needs small ($\sim$0.001) eccentricities and a massive 
belt which probably could not exist during the time needed for such accumulation. 
Therefore Ipatov (2001) considered that large TNOs and some main-belt asteroids
could be formed directly from dust rarefied condensations. 
It is assumed by many authors that a lot of dust condensations were 
formed from a dust disk around the forming Sun.
These initial condensations 
coagulated under collisions and formed larger condensations, which compressed 
and formed solid planetesimals. 
To our opinion, during the time needed for compression of 
condensations into planetesimals, some largest final condensations 
could reach such masses that they formed initial planetesimals with diameter 
equal to several hundreds kilometers.
As in the case of accumulation 
of planetesimals, there could be a "run-away" accretion of condensations. 
Some smaller objects (TNOs, 
planetesimals, asteroids) could be debris of larger objects, and other such 
objects could be formed directly by compression of condensations. 

\begin{figure}
\plottwo{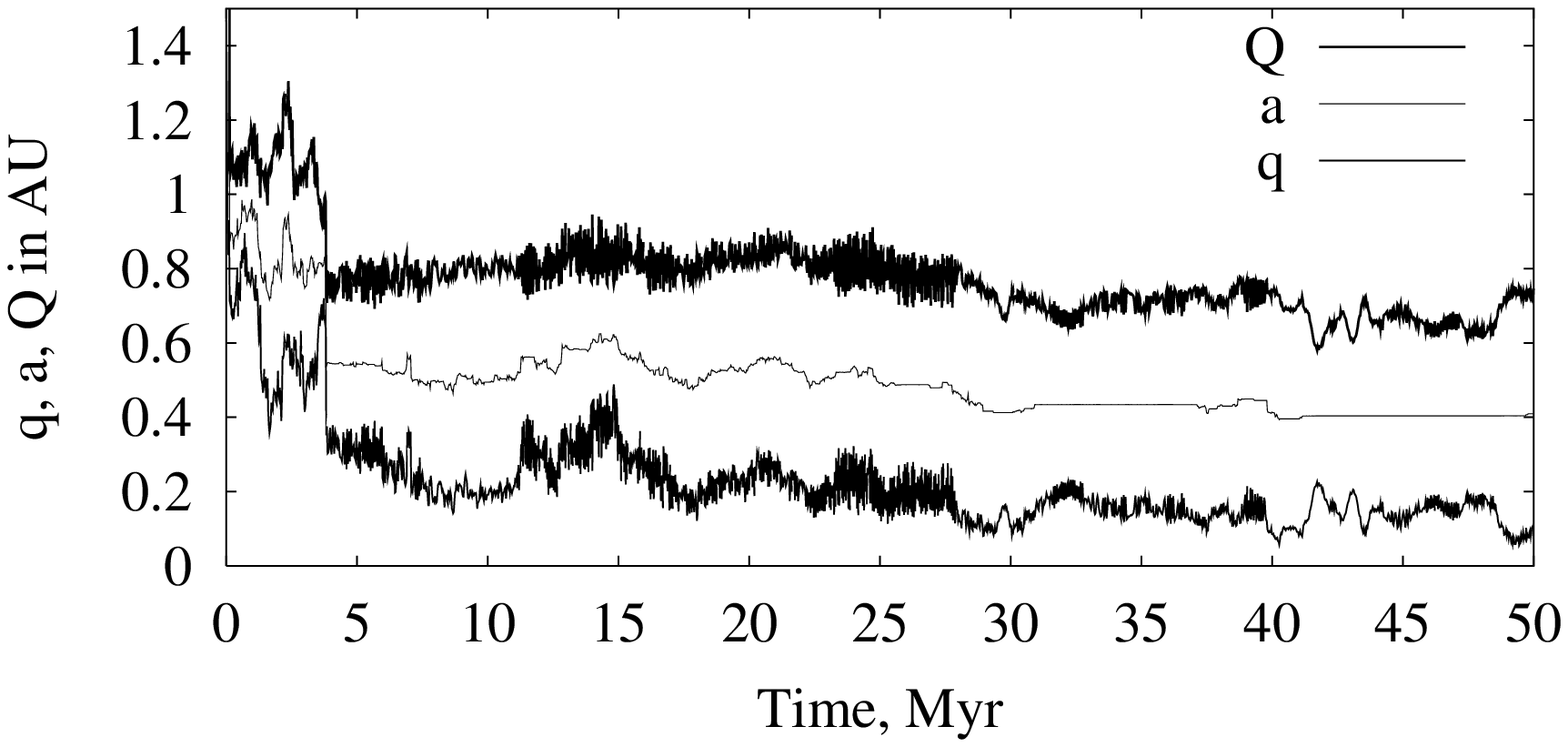}{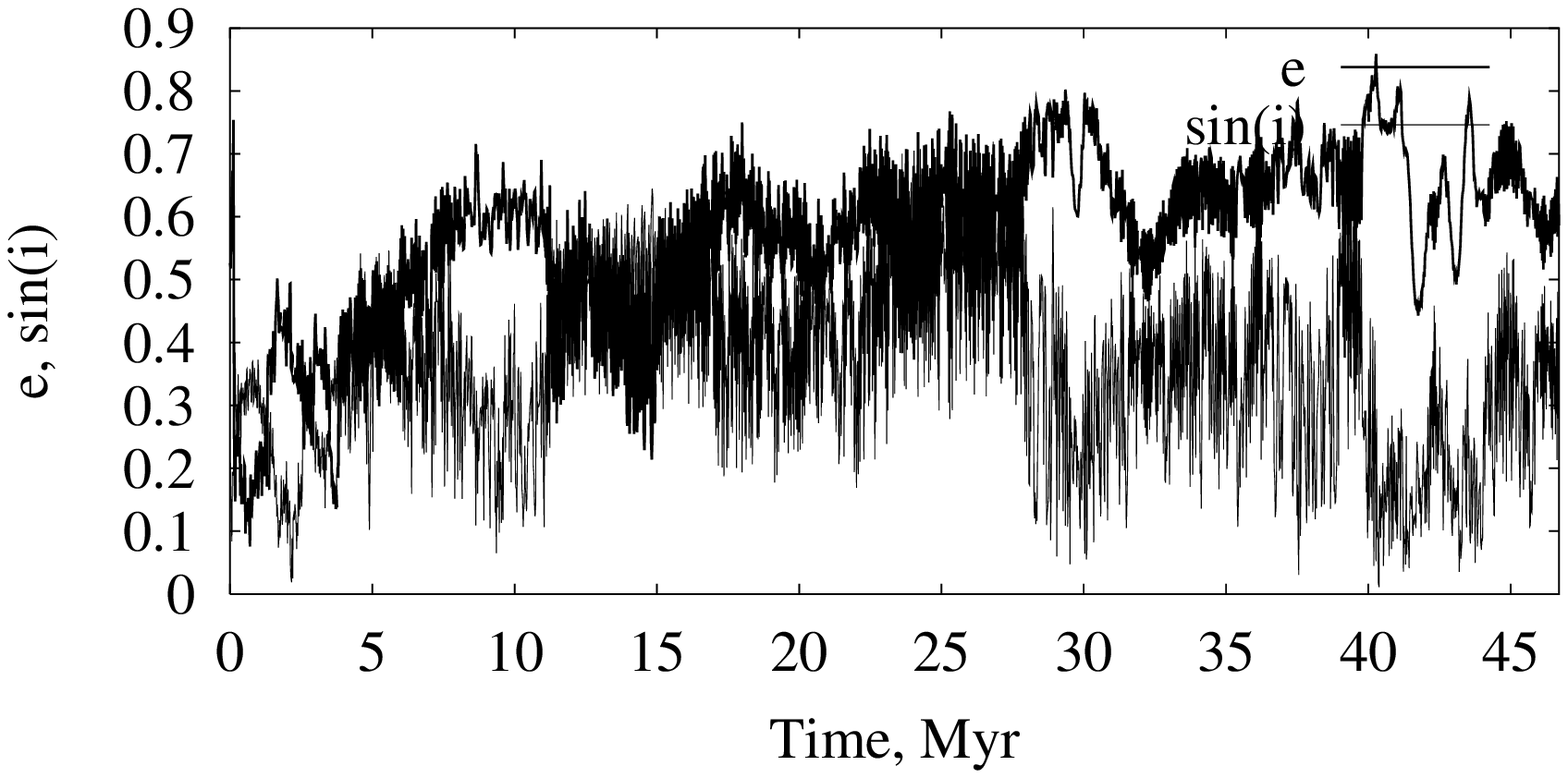}

\caption{Time variations in $a$, $q$, $Q$, $e$, sin$i$ for a former JCO in intial 
orbit close to that of Comet 10P  
($a$$>$1.5 AU at $t$$<$0.123 Myr) }

\end{figure}

     A small portion of planetesimals from the feeding zone of the giant 
planets that entered into the trans-Neptunian region could be left in 
eccentrical orbits beyond Neptune and became so called ''scattered disk 
objects'' (SDOs). The end of the bombardment of the terrestrial planets 
could be caused mainly by those planetesimals that had become SDOs. 
Our estimates (Ipatov 1995, 2001) showed that collisional lifetimes of 1-km TNOs 
and asteroids are about 1 Gyr. Typical TNOs can be even more often 
destroyed by SDOs than by other TNOs. 
Mutual gravitational interactions of TNOs 
can play a larger role in variations of their orbital elements than collisions.

\section{Orbital Evolution of Jupiter-Family Comets}

The motion of TNOs to Jupiter's orbit was investigated by several authors 
(e.g., Levison \& Duncan 1997). 
Proceeding from the total of $5\cdot10^9$ 1-km TNOs within 30$<$$a$$<$50 AU, assuming 
the mean time $T_{JCO}$$\approx$0.13 Myr for a body to move in a Jupiter-crossing orbit, 
and using the same formulas and other estimates as those in (Ipatov 2001), we obtain 
that about $10^4$ of former 1-km TNOs now are Jupiter-crossers.
% and $3\cdot10^3$ of them are Jupiter-family comets. 
%Of course, these estimates of Jupiter-crossers are very approximate. 
In the present paper we pay the main attention to the migration
of Jupiter-crossing objects (JCOs). 
Our present investigations are based on our runs of the orbital evolution of 
thousands JCOs 
%(Ipatov \& Hahn 2001 considered 48 orbits of JCOs)
under  gravitational influence of all 
planets, except for Mercury and Pluto, for intervals $T_S$$\ge$10 Myr 
(for Comet 2P we also considered Mercury). 
The integration package by Levison \& Duncan (1994) was used. 
Below we present the results obtained only by the method by Bulirsh and Stoer 
(BULSTO code), though we also considered the evolution of thousands of objects
using a symplectic method. In some cases in our runs (especially,  
when objects can get close to the Sun) we obtained a considerable difference 
between the results obtained by BULSTO and a symplectic method. % (RMVS3 code). 
For BULSTO the relative error per integration step was taken (depending on the run) 
to be less than $\varepsilon$ which was $10^{-9}$, $10^{-8}$, or some intermediate value.

\begin{figure}
\plottwo{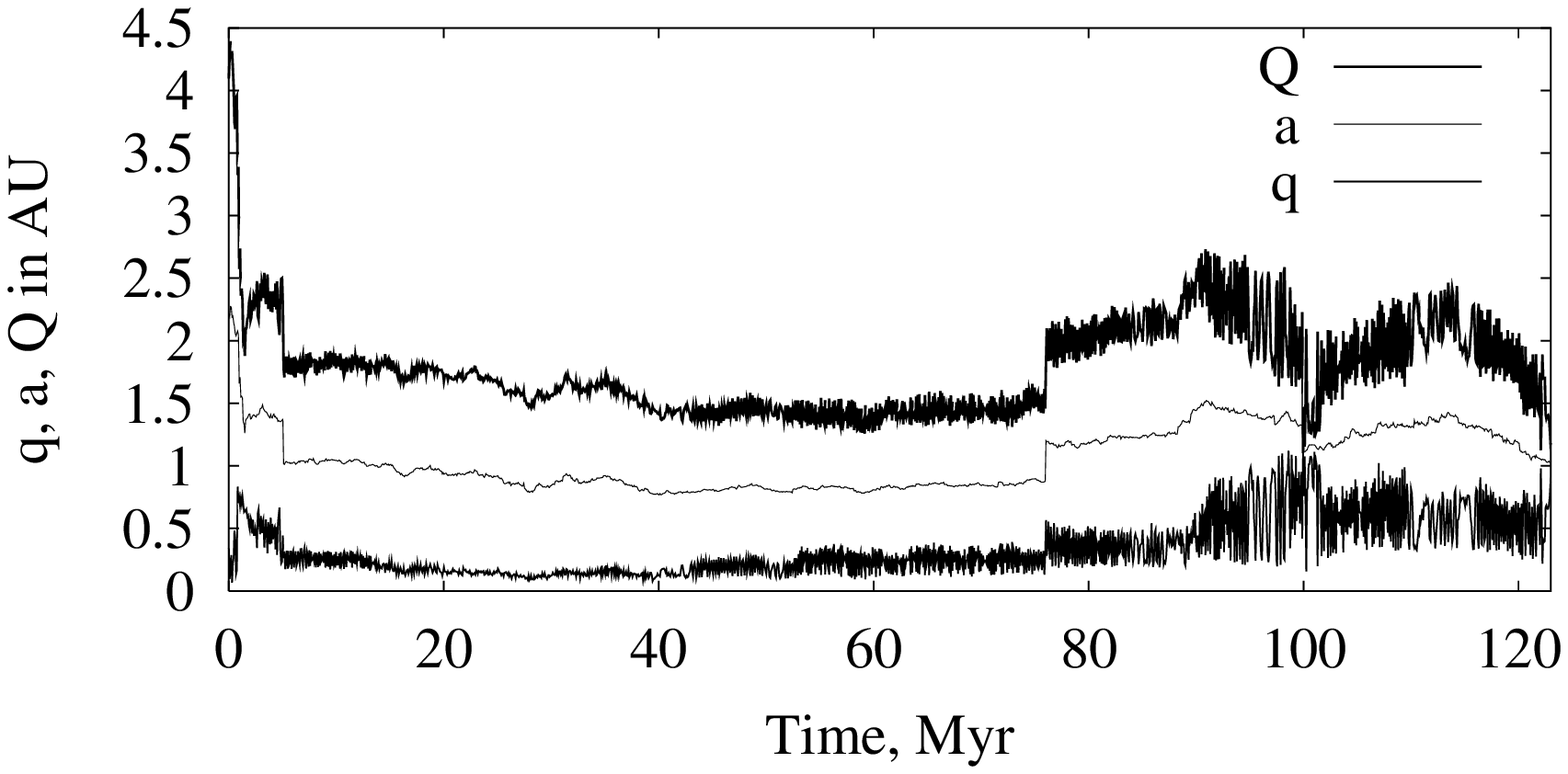}{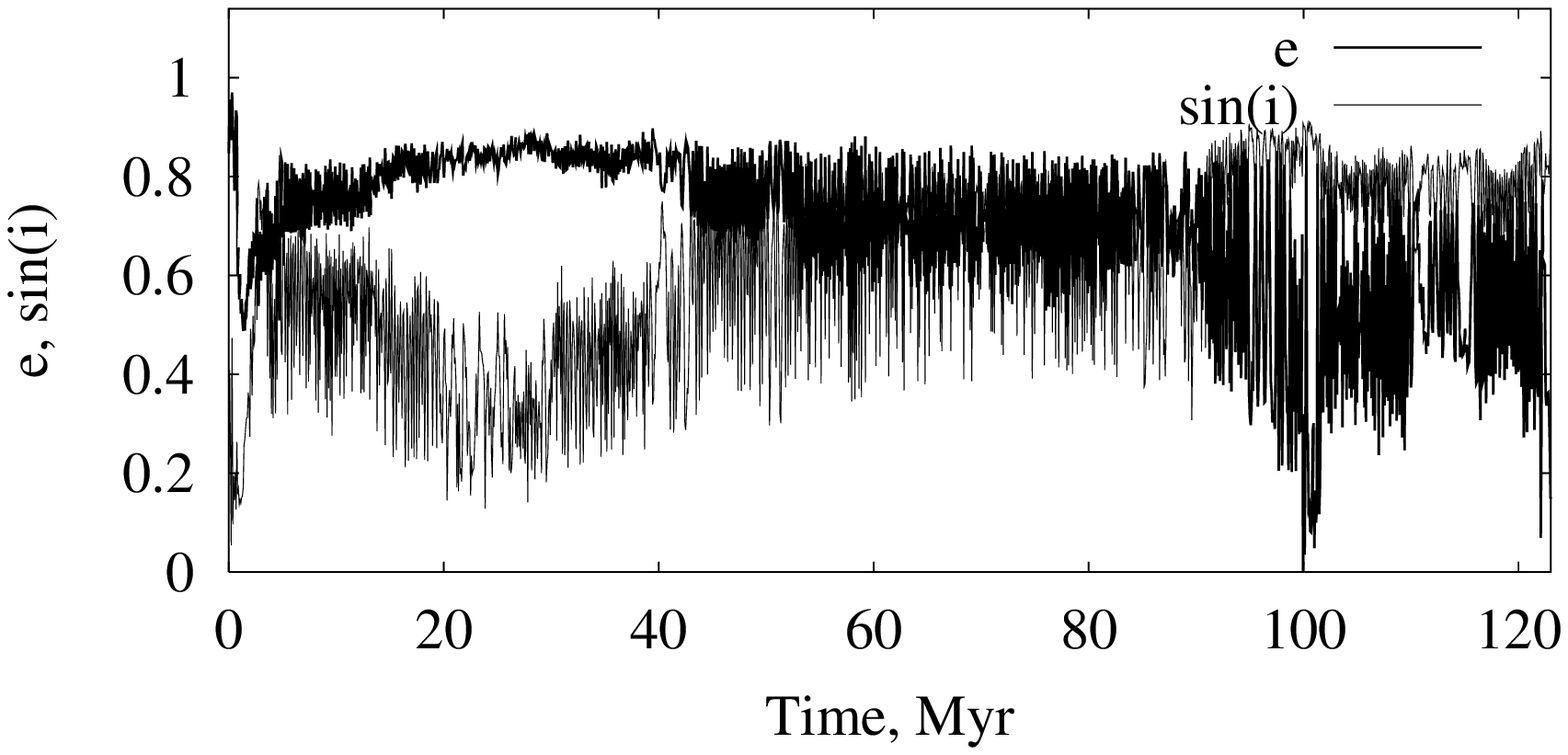}

\caption{Time variations in $a$, $q$, $Q$, $e$, sin$i$ for a former JCO in intial 
orbit close to that of Comet 2P  
}

\end{figure}
     In the first series of runs (denoted as $n1$) we investigated the orbital 
evolution of $N$=1900 JCOs moving in initial orbits close to those of 20 real 
JCOs with period 5$<$$P_a$$<$9 yr. In each of other series of runs we considered initial 
orbits close to that of one comet (2P, 9P, 10P, 22P, 28P, or 39P). 
We also studied the evolution of 
asteroids initially moving in the resonances 3:1 and 5:2 with Jupiter. 
Approximate values of initial semi-major axes $a$, eccentricities $e$ and 
inclinations $i$ of considered objects are presented in Table 1. 
For JCOs we varied only initial mean anomaly, 
and for asteroids we varied also initial value of the longitude of the 
ascending node. Examples of time variations in orbital elements are presented in 
Figs. 1-2. As these two objects have large probabilities of collisions with the 
terrestrial planets, they are not included in the Table. 
%In Fig. 3 during last 6 Myr the object  moved in orbit with $a$$\approx$1 AU.
%The distribution of migrating objects with their semimajor-axes is presenting in Fig. 3. 
In Fig. 3 we present the time in Myr during which 
7852 former JCOs had semi-major axes 
in the interval with a width of 0.005 AU (left) or 0.1 AU (right).

\begin{figure}

%\plottwo{a115aq2.eps}{a115ei2.eps}

\plottwo{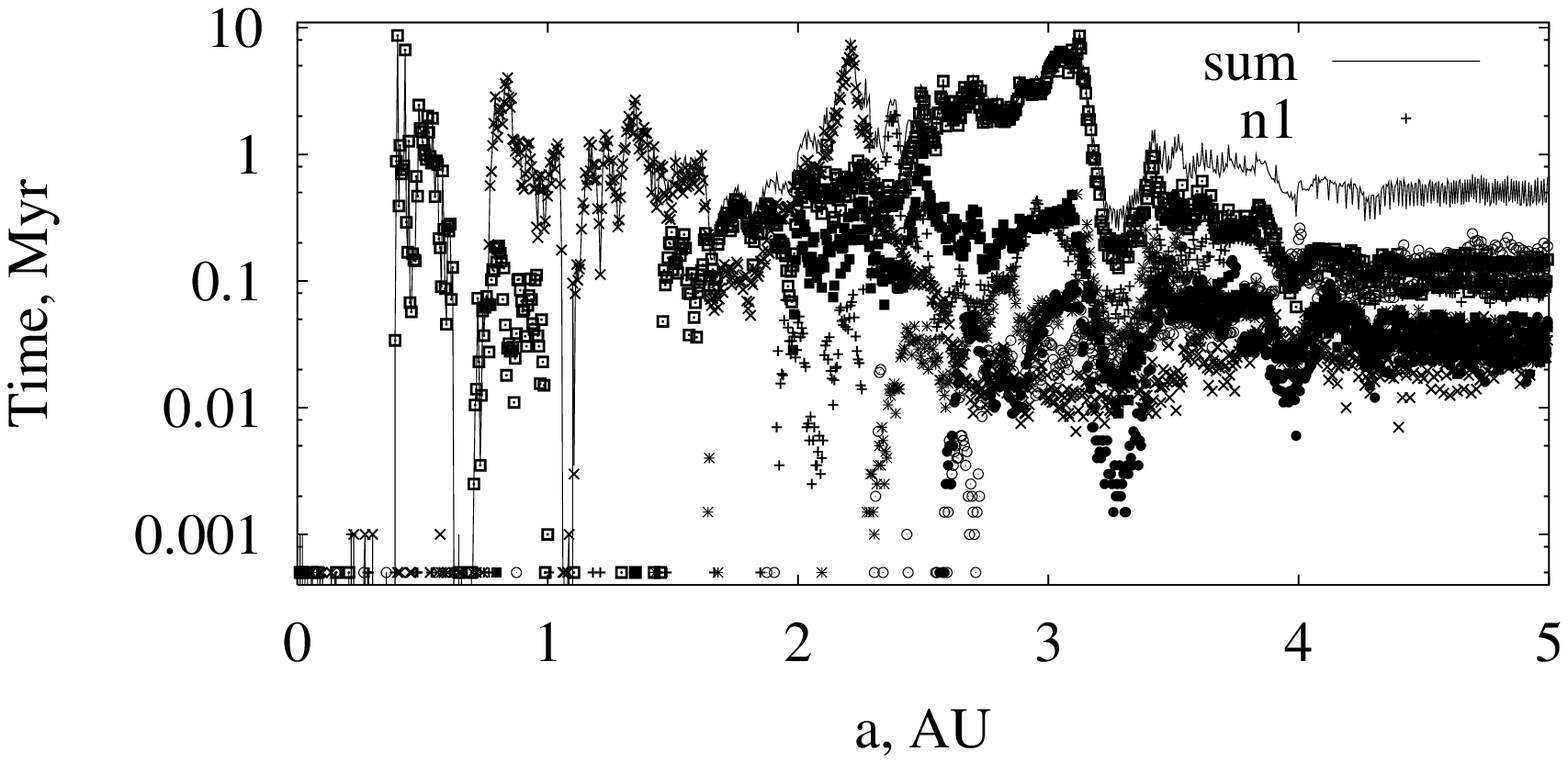}{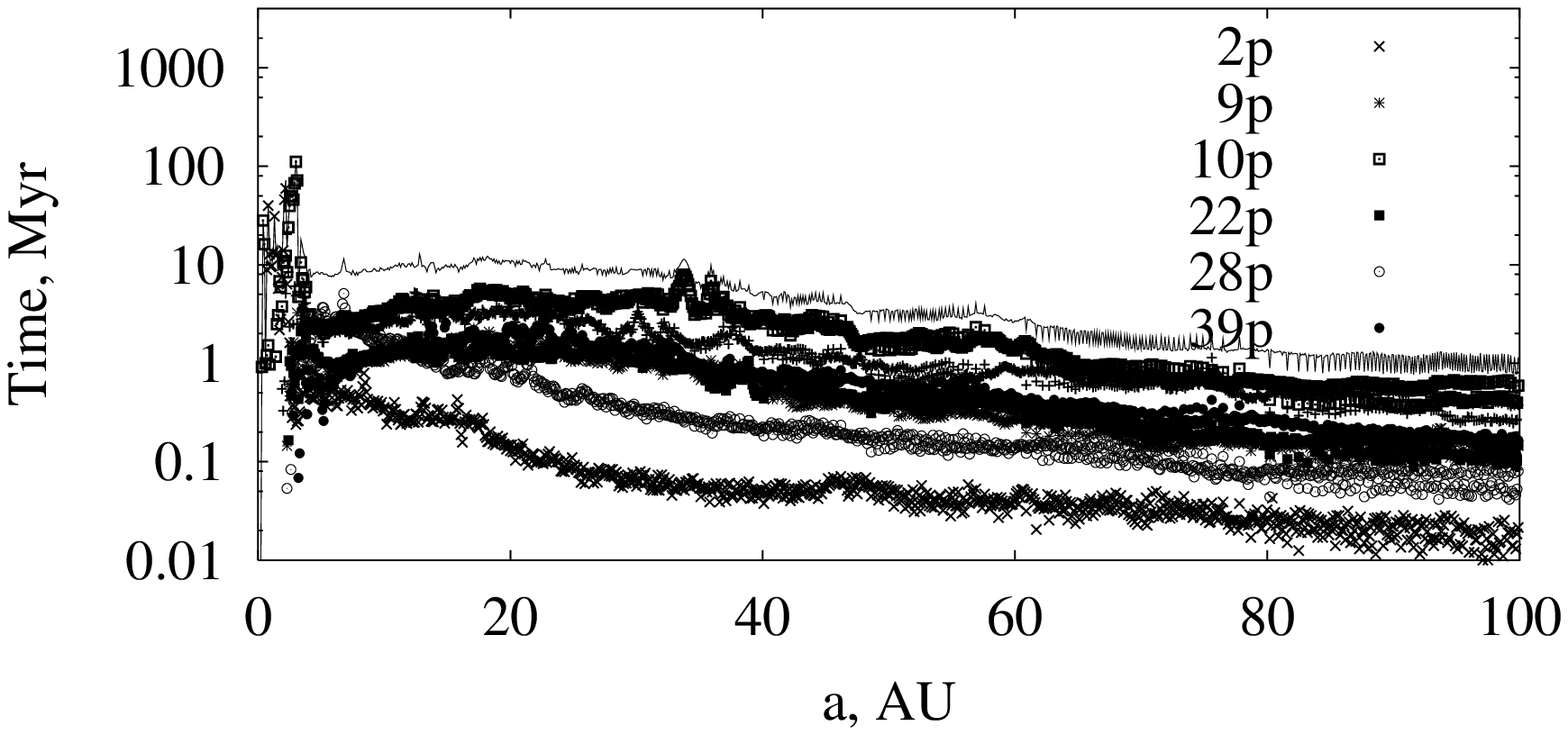}

\caption{Distribution of migrating objects with their semimajor-axes}

\end{figure}

We did not simulate collisions of objects with planets, but
basing on orbital elements obtained with a step 500 yr, for all $N$ objects we 
calculated the total impacting probability $P_\Sigma$ and the total time interval 
$T_\Sigma$ to reach  perihelion distance
$q$ less than a semi-major axis of a planet,  and then 
impact probabilities per one object $P$=$10^{-6}P_r$=$P_\Sigma/N$ and 
$T_r$=$T_\Sigma/N$ during $T_S$ were estimated. 

\begin{table}
\caption{Values of $T$ (in Kyr), $P_r$, and $r$ obtained by the BULSTO code 
(Venus=V, Earth=E, Venus=V)}

$ \begin{array}{lllllccccccc} 

\hline	

  & & &&&$V$ & $V$ & $E$ & $E$ & $M$ & $M$ & \\

\cline{6-11}

 & N& a& e&i&P_r & T & P_r & T & P_r & T & r\\

\hline

n1&1900& &&& 2.42 & 4.23 & 4.51 & 7.94 &  6.15 & 30.0 & 0.7 \\
$2P$ & 501 &2.22 &0.85&12&226 &504  & 162 & 548 &  69.4  & 579 & 19. \\
$9P$ & 800&3.12 &0.52&10&1.34 & 1.76 & 3.72 & 4.11 &     0.71 & 9.73 & 1.2 \\
$10P$ & 2149&3.10&0.53&12&28.3 & 41.3 & 35.6 & 71.0 &   10.3 & 169. & 1.6 \\
$22P$ & 1000&3.47 &0.54&4.7&1.44&2.98&1.76 & 4.87 &     0.74 & 11.0 & 1.6 \\
$28P$ & 750 &6.91&0.78&14& 1.7 & 21.8 & 1.9 & 34.7 &     0.44 & 68.9 & 1.9 \\
$39P$ & 750 &7.25&0.25&1.9& 1.06&1.72&  1.19& 3.03 & 0.31 & 6.82 & 1.6 \\
$total$& 7850&&&& 17.9 & 37.7 & 18.8 & 51.5 &  5.29 & 85.9 & 2.6 \\
3:1& 288 & 2.5&0.15&10& 940 &1143& 1223& 1886 &       371& 3053 & 2.3 \\
5:2& 288 & 2.82&0.15&10& 95.8 &170 & 160 & 304 &       53.7 & 780 & 1.0 \\

\hline
\end{array} $ 
\end{table}

In Table 1 we also present
the ratio $r$ of time spent in Apollo orbits
($a$$>$1 AU, $q$=$a(1-e)$$<$1.017 AU) at $e$$<$0.999 to that in Amor orbits 
(1.017$<$$q$$<$1.33 AU). 
  One object initially moving
in orbit close to that of Comet 10P after having  Aten-type orbit 
($a$$<$1 AU, $Q$=$a(1+e)$$>$0.983 AU) during 3 Myr 
(the probability of its collision with the Earth was 0.34)
got orbits with aphelion distance $Q$$<$0.983 AU (Fig. 1). The probability of its collision
with Venus during $t$$\le$50 Myr
was $\approx$3, so more probable that it collided Venus at $t$$\approx$15 Myr.
%The total probabilities of collisions with Venus and Earth for all
%other 7851 objects were 0.18 and 0.17, respectively.

Specific mass of the matter delivered by JCOs to an inner planet (normalized to 
its mass) turns out to be nearly the same for Earth and Venus though greater for Mars.
The total collisional probability with the inner planets was mainly caused 
by a small ($\sim$0.01-0.001) fraction of bodies residing in orbits deep inside Jupiter's 
orbit for more than 1 Myr. Fifteen considered objects with initial orbits
close to those of 10P and 2P 
moved in Earth-crossing orbits with 1$<$$a$$<$2 AU during
more than 0.5 Myr each. At $n1$, objects moved with periods $P_a$$<$10 
and 10$<$$P_a$$<$20 yr during 11\% and 21\% of $T_{JCO}$,
%their motion in Jupiter-crossing orbits, 
respectively. 
%Roughly 1 of 300 JCOs collides with the Sun.
More details of the runs can be found in our papers 
in htpp://arXiv.org/archive/astro-ph.

\section{Migration of Trans-Neptunian Objects to the Near-Earth Space}

In total, 6852 considered JCOs moved during more than 160, 470, 65, 140, and 10 Myr in
Amor, Apollo, and Aten orbits, 
orbits with 1$<$$a$$<$2 AU and $q$$<$1 AU, and orbits with $Q$$<$0.983, respectively. 
So, if we consider $10^4$ former 1-km TNOs
now moving in Jupiter-crosing orbits, then for $T_{JCO}$=0.13 Myr
their number for the above 
five types of orbits is about 1600, 4700, 650, 1400, and 100,
respectively. As we simulated mainly orbits with large probabilities of
collisions with the Earth, the above numbers can be smaller by a factor
of several. Mean
eccentricities of such orbits are larger and the probabilities of
collisions with the terrestrial planets are smaller than 
those of the observed NEOs. Probably, most of such former TNOs (extinct 
comets) are not yet observed, as most of the time they move relatively 
far from the Earth. 
It is considered that about 750 of 1-km 
bodies are located in the Earth-crossing orbits (half of them are 
in orbits with $a$$<$2 AU), although this number does not include those 
in high eccentric orbits. 
 Our estimates 
show that, in principle, the trans-Neptunian belt can provide a considerable 
portion of Earth-crossing objects, at least many of those with $a$$>$2 AU, but, 
of course, some NEOs came from the main asteroid belt.
It may be possible to explore former TNOs near the Earth's orbit without 
sending spacecrafts beyond Neptune.
     Based  on the estimated collision probability $P$=$6\cdot10^{-6}$ (this 
value is a little larger than that for $n1$, but is smaller by an order of 
magnitude than $P$=$7\cdot 10^{-5}$ obtained for 7852 JCOs) and assuming 
the total mass of planetesimals 
that ever crossed Jupiter's orbit is $\sim$100$m_\oplus$ 
($m_\oplus$ is mass of the Earth), we found 
that the total mass of bodies impacted on the Earth is 
$6\cdot10^{-4}m_\oplus$. If ices composed 
only a half of this mass, then the total mass of ices that were delivered to 
the Earth from the feeding zone of the giant planets turns out by the factor 
of 1.5 greater than the mass of the Earth oceans. Ancient oceans on Mars and Venus
could be also large. 

This work was supported by NRC (0158730), NASA (NAG5-10776), INTAS (00-240), 
and RFBR (01-02-17540). 
%The author is thankful to V.I. Ipatova for the help in calculations.

\end{document}